\documentclass[12pt]{article} 
\usepackage[sectionbib]{natbib}
\usepackage{array,epsfig,fancyheadings,rotating}
\usepackage[]{hyperref}  

\usepackage{sectsty, secdot}
\sectionfont{\fontsize{12}{14pt plus.8pt minus .6pt}\selectfont}
\renewcommand{\theequation}{\thesection\arabic{equation}}
\subsectionfont{\fontsize{12}{14pt plus.8pt minus .6pt}\selectfont}

\usepackage{amsmath}
\usepackage{amssymb}
\usepackage{amsfonts}
\usepackage{multirow}
\usepackage{amsthm}
\usepackage{color}

\newtheorem{theorem}{Theorem}

\theoremstyle{definition}

\newcommand{\bX}{\boldsymbol{X}}
\newcommand{\bx}{\boldsymbol{x}}

\newcommand{\bU}{\boldsymbol{U}}

\newcommand{\bgamma}{\boldsymbol{\gamma}}
\newcommand{\blambda}{\boldsymbol{\lambda}}
\newcommand{\bGamma}{\boldsymbol{\Gamma}}
\newcommand{\bbeta}{\boldsymbol{\beta}}
\newcommand{\bdelta}{\boldsymbol{\delta}}
\newcommand{\boldeta}{\boldsymbol{\eta}}
\newcommand{\bzero}{\boldsymbol{0}}
\newcommand{\tX}{\widetilde{\bX}}
\newcommand{\tW}{\widetilde{W}}
\newcommand{\hW}{\widehat{W}}
\newcommand{\Var}{{\rm Var}}

\newcommand{\Cov}{{\rm Cov}}

\newcommand{\Corr}{{\rm Corr}}

\newcommand{\T}{\mathsf{T}}
\newcommand{\bbP}{\mathbb{P}}
\newcommand{\bbG}{\mathbb{G}}

\begin{document}


\renewcommand{\baselinestretch}{2}
\markright{ \hbox{\footnotesize\rm Statistica Sinica
}\hfill\\[-13pt]
\hbox{\footnotesize\rm
}\hfill }

\markboth{\hfill{\footnotesize\rm MIN QIAN, BIBHAS CHAKRABORTY, RAJU MAITI and YING KUEN CHEUNG} \hfill}
{\hfill {\footnotesize\rm SEQUENTIAL TEST OF INTERACTION EFFECTS} \hfill}

\renewcommand{\thefootnote}{}
$\ $\par


\fontsize{12}{14pt plus.8pt minus .6pt}\selectfont \vspace{0.8pc}
\centerline{\large\bf A SEQUENTIAL SIGNIFICANCE TEST FOR }
\vspace{2pt} \centerline{\large\bf TREATMENT BY COVARIATE INTERACTIONS}
\vspace{.4cm} \centerline{Min Qian$^1$, Bibhas Chakraborty$^{2,3}$, Raju Maiti$^2$ and Ying Kuen Cheung$^1$} \vspace{.4cm} \centerline{\it
$^1$Columbia University, $^2$National University of Singapore and $^3$Duke University} \vspace{.55cm} \fontsize{9}{11.5pt plus.8pt minus
.6pt}\selectfont


\begin{quotation}
\noindent {\it Abstract:}
Biomedical and clinical research is gradually shifting from the traditional ``one-size-fits-all'' approach to the new paradigm of personalized medicine. An important step in this direction is to identify the treatment by covariate interactions. We consider the setting in which there are potentially a large number of covariates of interest. Although a number of novel machine learning methodologies have been developed in recent years to aid in treatment selection in this setting, few, if any, have adopted formal hypothesis testing procedures. In this article, we present a novel testing procedure based on $m$-out-of-$n$ bootstrap that can be used to sequentially identify variables that interact with treatment.  We study the theoretical properties of the method and show that it is more effective in controlling the type I error rate and achieving a satisfactory power as compared to competing methods, via extensive simulations. Furthermore, the usefulness of the proposed method is illustrated using real data examples, both from a randomized trial and from an observational study.

\vspace{9pt}
\noindent {\it Key words and phrases:}
Double robustness; Forward stepwise testing; 
$m$-out-of-$n$ bootstrap; Non-regular asymptotics; Personalized medicine.
\par
\end{quotation}\par

\def\thefigure{\arabic{figure}}
\def\thetable{\arabic{table}}

\renewcommand{\theequation}{\thesection.\arabic{equation}}

\fontsize{12}{14pt plus.8pt minus .6pt}\selectfont

\setcounter{section}{1} 
\setcounter{equation}{0} 

\lhead[\footnotesize\thepage\fancyplain{}\leftmark]{}\rhead[]{\fancyplain{}
	\rightmark\footnotesize\thepage}

\noindent {\bf 1. Introduction}

Due to patient heterogeneity in response to various aspects of any treatment program, biomedical and clinical research is gradually shifting from the traditional ``one-size-fits-all'' treatment settings to the cutting-edge paradigm of personalized medicine. An important step in this direction is to identify the treatment by covariate interactions.  In the conventional approach, investigators would first identify a set of key covariates. Treatment by covariate interactions would then be examined either by comparing treatment vs. control in subgroups defined by the key covariates, or by  testing the regression coefficients of the treatment by covariate interaction terms in a multivariable linear model. These approaches, however, are either infeasible or unreliable due to overfitting, when there is a moderate to large number of covariates under consideration.

In recent years, a number of novel methodologies have been developed to identify treatment by covariate interaction effects based on a large set of covariates, with the ultimate goal of optimizing treatment selection. This includes ranking methods (\citealt*{gunter2011, tian2011}; \citealt{chen2017}), regression based methods (\citealt*{qian2011, lu2013}; \citealt{tian2014}; \citealt*{fan2016}), weighted classification type learning methods (\citealt*{Orellana10}, \citealt{zhang2012a, zhao2012, huang2014,  liu2018}), tree based methods (\citealt{su2008, laber2015, tsai2016}), and functional data approaches (\citealt{mckeague2014, adam2015, laber2018}),  to mention a few. 
A major
gap in this literature, however, has been the scarcity of formal hypothesis testing
procedures that take variable selection into account.  

There is a limited number of papers that discussed novel hypothesis testing approaches for subgroup identification of enhanced treatment effect.   \citet*{shen2015inference} developed a likelihood-based test for the existence of a subgroup based on linear logistic-normal mixture models. \citet*{fan2017} proposed a method to test and identify a subgroup using change-point techniques.  \citet*{wager2018} investigated a forest-based method for treatment
	effect estimation and inference. \cite*{shi2019} proposed a nonparametric test to assess the incremental value of a given set of new variables in optimal treatment decision making conditional on an existing set of prescriptive variables. These methods focused on the test of non-linear treatment effects, and worked well with a relatively small set of covariates. Although \cite*{shi2019} considered application of their method in a forward stepwise fashion and studied its variable selection properties, the derived $p$-value loses its interpretation when used with forward selection.
	In the presence of large  number of covariates, \cite{shen2016} proposed a kernel based method to identify whether there is interaction between the treatment and a group of covariates; however, this is applicable only to a randomized trial setting. \citet*{zhao2017selective} considered a lasso based inference  approach to identify informative covariates under the condition that both the propensity score and the main effect of covariates on outcome are well estimated.
	
In a methodologically analogous setting, gene-environment interaction has been extensively studied in the field of genetics. The most common approach is to test the interaction of each genetic marker and an environmental exposure separately, and then adjust for multiple comparison. To improve power and to reduce the burden of multiple comparison, several global tests have been proposed to assess the joint interaction effect between a marker set and an environmental variable (e.g. \citealt{lin2013test, marcea2015}). The validity of these methods, however, rely on approximately correct model of the main effects of markers and environmental factors. 

In the current paper, we consider data from either randomized trials or observational studies. We aim to identify covariates that interact with treatment, among a large set of candidate covariates, via a sequential testing procedure. First, a marginal screening test is used to detect whether there is any covariate that significantly interacts with treatment. If the test is significant, then we proceed to test whether there are additional treatment by covariate interactions in a forward stepwise fashion. The procedure continues until the p-value exceeds the pre-specified level of significance. 
Forward stepwise regression has been extensively studied (\citealt*{barron2008, donoho2006, wang2009, ing2011}). However, the vast majority of the literature focused on studying the variable selection consistency properties, instead of hypothesis testing. In real applications, to enhance reproducibility, clinicians would  often like to have
 some inferential guarantees for the selected covariates, and variable selection consistency theory alone would not satisfy their needs. This type of selective inference problems have drawn a lot of attention  from statisticians, and various methods have been proposed in the prediction literature (e.g., \citealt{buhlmann2013, zz2014, lockhart2014,vandegeer2014,mckeague2015, ning2017, alex2018}).  




We propose to calibrate our test statistic either by directly sampling from the null (if it is estimable) or by using the $m$-out-of-$n$ bootstrap. 
The $m$-out-of-$n$ bootstrap is a general tool for conducting valid statistical inference for non-regular parameters (\citealt*{shao1994,bickel1997}). 
It is the usual nonparametric bootstrap (\citealt*{efron1979}) except that the resample size $m$ is of a smaller order than the original sample size $n$. With an appropriate choice of $m$, the $m$-out-of-$n$ bootstrap acts as a smoothing operation on the empirical distribution of the data. 
Data-driven methods for choosing $m$ in various contexts were proposed in \citet*{hall1995}, \citet*{lee1999}, \citet*{bickel2008} and \citet*{chakraborty2013}. In the current paper, an adaptive choice of $m$ 
is developed. 
The proposed test is valid as long as either the propensity score is known (or modeled correctly) or the main effects model of covariates on outcome is correctly specified.

The paper is organized as follows. In Section \ref{sec:marginal}, we set up the framework and describe the initial marginal screening test for identifying the variable that most strongly interacts with treatment in the randomized trial setting. In Section \ref{sec:sequential}, we present the sequential test procedure. In Section \ref{sec:observe}, we extend our method to allow for double robustness, which is particularly useful in observational studies where the propensity score model is unknown. 
In Section \ref{sec:numerical}, we conduct simulations comparing proposed methods with existing competitors, and illustrate the methods using two data examples. 
We conclude with a discussion in Section \ref{sec:discussion}.  
Proofs of the theorems and details of simulations are presented in the Supplementary Materials. 

\setcounter{equation}{0} 
\section{Marginal Screening Test for Randomized Trials}
\label{sec:marginal}

\subsection{Marginal regression}

Suppose we are given pre-treatment information, treatment assignment, and outcomes from $n$
patients. Further suppose there are only two competing treatments $A\in\{0,1\}$. Let $\bX\in\mathbb{R}^p$ be the vector of pre-treatment variables, and $Y$ be a scalar outcome. Let
$q_0(\bx) := P(A=1|\bX=\bx)$ be  the propensity of receiving
treatment $1$ in the observed data as a function of pre-treatment variables $\bX=\bx$.  In this section and Section \ref{sec:sequential}, we assume that the data come from a randomized trial, so $q_0(\bx)$ is known. 

We frame the problem in terms of the model
\begin{equation}
Y= h_0(\bX) + (\alpha_0 + \bX^\T\bbeta_0)A  + \epsilon, \label{eqn:simplemodel}
\end{equation}
where $\bbeta_0\in\mathbb{R}^{p}$, $h_0(\bX) := E(Y|\bX, A=0)$, and the error term $\epsilon$ has mean zero, finite variance, and is uncorrelated with $A-q_0(\bX)$ and $(A-q_0(\bX))\bX$. The term $\alpha_0 + \bX^\T\bbeta_0$ models $T(\bX)  := E(Y|\bX, A=1)- E(Y|\bX, A=0)$, the causal treatment effect for patients
with pre-treatment information $\bX$; thus, $(\alpha_0 + \bX^\T\bbeta_0)A$ is the treatment-by-covariate interaction model.
In this paper, we propose a sequential testing procedure to identify covariates that interact with treatment. 

As an initial step, we are interested in testing whether there is any treatment-by-covariate interaction. That is, we want to test
\begin{align}
H_0: \bbeta_0 =\bzero  \mbox{ vs. } H_a: \bbeta_0\neq \bzero. \label{eqn:h0}
\end{align}
Our proposed method is based on fitting $p$ working marginal regression models, and conducting a single test on the marginal regression coefficient of the most informative predictor of the causal treatment effect $T(\bX)$.   Specifically,  we can write $E(Y|\bX,A) = E(Y|\bX) + T(\bX)W$, where $W := A-q_0(\bX)$ (see \citealt{robins1994}). 
For $k=1,\ldots,p$, consider the working model
$T(\bX) = \alpha_k + \theta_kX_k$.  The $k$th marginal regression model aims to estimate
\begin{align}
(\alpha_k,\theta_k) = \arg\min_{(\alpha,\theta)}E\left\{\big[Y - E(Y|\bX) - (\alpha + \theta X_k)W\big]^2\right\}. \label{eqn:sol1}
\end{align}
And the index of the most informative predictor of $T(\bX)$ is
\begin{align*}
k_0 = &\arg\min_kE\left\{\big[Y - E(Y|\bX) - (\alpha_k + \theta_k X_k)W\big]^2\right\}.
\end{align*}

By taking first order derivative of (\ref{eqn:sol1}) with respect to $(\alpha,\theta)$ and noting that $E(W|\bX)=0$, we have under model (\ref{eqn:simplemodel}),
\begin{align*}
\theta_k = &\, \frac{E\big[W(Y- E(Y|\bX))X_k'\big]}{E(WX'_k)^2}
= \, \frac{\Cov(WX'_k, W\bX^T\bbeta_0)}{E(WX'_k)^2}\\
\mbox{ and }\quad k_0 = &\,\arg\max_k \big|\Corr(WX'_k, W\bX^T\bbeta_0)\big|,
\end{align*}
where  $X'_k = X_k - E(W^2X_k)/EW^2$. 
Assume that $k_0$ is unique when $\bbeta_0\neq\bzero$. We can verify that $\bbeta_0=\bzero$ if and only if $\theta_k=0$ for all $k=1,\ldots,p$. Thus hypothesis (\ref{eqn:h0}) is equivalent to
\begin{align}
H_0: \theta_0 =0 \mbox{ vs. } H_a: \theta_0\neq 0, \label{eqn:h0theta}
\end{align}
where $\theta_0 = \theta_{k_0}$. Using randomized trial data,
we can estimate $k_0$ and $\theta_0$ by
\begin{align*}
\hat k_n = &\arg\min_{k \in\{1,\ldots,p\}}\bbP_n\left\{\Big[Y - \hat\phi_n(\bX) - (\hat\alpha_k + \hat\theta_k X_k)W\Big]^2\right\}\\
\mbox{and } \quad \hat\theta_n := &\;\hat\theta_{\hat k_n} = \left\{\bbP_n\left[\big(W\hat X'_{\hat k_n}\big)^2\right]\right\}^{-1}\bbP_n\Big[W(Y-\hat\phi_n(\bX))\hat X'_{\hat k_n}\Big],
\end{align*} 
respectively, where $\bbP_n$ denotes the sample average, $\hat\phi_n(\bX)$ is an estimate of $E(Y|\bX)$,  $(\hat{\alpha}_k,\hat{\theta}_k) = \arg\min_{(\alpha,\theta)}\Big\{\bbP_n\big[Y - \hat\phi_n(\bX) - (\alpha + \theta X_k)W\big]^2\Big\}$,
and $\hat X'_k = X_k-\bbP_n(W^2X_k)/\bbP_nW^2$.

Our first result gives the asymptotic distribution of $\hat\theta_n$.
\begin{theorem}
	\label{thm:simple}	Assume i) $EX_k^4 < \infty$  for $k=1,\ldots,p$; ii) the error term 
	$\epsilon$ in model (\ref{eqn:simplemodel}) has mean zero, finite variance and is uncorrelated with $W$ and $W\bX$; and iii) $\hat{\phi}_n(\bX)$ is estimated from a $P$-Donsker class of measurable functions, and
	there exists some $\tilde\phi(\bX)$ such that $E[\hat\phi_n(\bX)- \tilde \phi(\bX)]^4\stackrel{P}{\to} 0$ for some  $\tilde\phi(\bX)$ that is fourth moment integrable as $n\to\infty$.  
	Suppose $k_0$ is unique when $\boldsymbol\beta_{0} \neq \bzero$. Then under model (\ref{eqn:simplemodel}),
	$$n^{1/2}(\hat\theta_n-\theta_0)\stackrel{d}{\to} \frac{Z_{k_0}}{E[WX'_{k_0}]^2}1_{\bbeta_{0}\neq \boldsymbol 0} + \frac{Z_{K}}{E[WX'_{K}]^2}1_{\bbeta_{0} =  \boldsymbol 0},$$
	where  $X'_k = X_k - E(W^2X_k)/EW^2$, $K=\arg\max_{k=1,\ldots,p}Z_k^2/E(WX'_{k})^2$,
	and  $(Z_1,\ldots,Z_p)^\T$ is a mean-zero normal random vector with covariance matrix
	$\Sigma$ given by that of the   random vector with components
	\begin{align*}
	WX'_{k}\bigg\{ Y-\tilde{\phi}(\bX)-\frac{E[W(Y-\tilde{\phi}(\bX))]}{EW^2}W 
	-\frac{E[WX'_{k}(Y-\tilde{\phi}(\bX))]}{E(WX'_{k})^2}WX'_{k}
	\bigg\}
	\end{align*}
	for $k=1,\ldots,p$, and $\Sigma$ is assumed to exist.
\end{theorem}
\noindent\textbf{Remark 1.} The uniqueness of $k_0$ is  assumed in order to make sure that the parameter $\theta_0$ is well defined under $H_a$. Note that for hypothesis testing purpose, the test is always calibrated  under the null distribution, which does not rely on this condition. Nonetheless,
this condition can be removed with a slight modification of hypothesis (\ref{eqn:h0theta}) and the test statistic. 
A modified version of Theorem \ref{thm:simple} without the uniqueness condition is presented in Section S3 of the Supplementary Materials. 

\noindent\textbf{Remark 2.}
$\hat\phi_n(\bX)$ is an estimate of $E(Y|\bX)$, which can be obtained via sample mean of $Y$, (regularized) regression from a linear model, or other non-parametric methods based on a Donsker class of functions that guarantee the convergence of $\hat\phi_n(\bX)$ to some function $\tilde{\phi}(\bX)$. 
	The result does not require $\hat\phi_n(\bX)$ to be a consistent estimate of $E(Y|\bX)$. However, a good estimate of $E(Y|\bX)$ may help reduce the variance of each $Z_k$, and thus $\hat{\theta}_n$. For example, assume $q_0(\bX)=1/2$. We can verify that
	\begin{align*}
	\Var (Z_k)
	= &\, \Var\left(WX'_{k}\big[ E(Y|X)-\tilde{\phi}(\bX)\big]\right) +
	\Var\left(WX'_{k}\epsilon\right) \\  
	& + \Var\left(W^2X'_{k}\Big[ (\bX-E\bX)^\T\bbeta_0 
	-\theta_kX'_{k}
	\Big]\right).
	\end{align*}
	So the variance of each $Z_k$ is minimized when $\tilde{\phi}(\bX) = E(Y|\bX)$.

One way to test $\theta_0=0$ is to estimate the null distribution of $n^{1/2}\hat{\theta}_n$ by setting $\bbeta_0=\bzero$ in Theorem  \ref{thm:simple} and replacing $\tilde{\phi}(\bX)$ with $\hat\phi_n(\bX)$ and expectation with sample average, respectively. The p-value can be calculated by comparing the observed test statistic $n^{1/2}\hat{\theta}_n$ with the estimated null distribution. 
Alternatively, below we introduce an $m$-out-of-$n$ bootstrap method to estimate the asymptotic distribution of $\hat{\theta}_n$. This method can then be easily extended to cases where the null distribution is difficult to estimate (see Section \ref{sec:observe}).

\subsection{The $m$-out-of-$n$ bootstrap}
 The $m$-out-of-$n$ bootstrap is a general tool to remedy bootstrap inconsistency due to nonsmoothness. 
	When the resample size $m$ is of a smaller order than $n$,
	the empirical distribution converges to the true  distribution
	at a faster rate than the analogous convergence of the
	$m$-out-of-$n$ bootstrap sample empirical distribution to the empirical distribution.
	Intuitively, this implies that the empirical distribution converges to the true distribution first and thus bootstrap resamples behave as if they were
	drawn from the true distribution.

For a selected resample size $m$, let  $\hat{\theta}_m^{*}$ be the analog of $\hat{\theta}_n$ based on the bootstrap sample of size $m$. Theorem  \ref{thm:bootstrap1} below shows the bootstrap consistency results. And the p-value can be calculated by comparing $n^{1/2}\hat{\theta}_n$ with the distribution of $m^{1/2}(\hat{\theta}_m^{*}-\hat{\theta}_n)$.


\begin{theorem}
	Assume all conditions in Theorem  \ref{thm:simple} hold. Suppose $m/n=O(1)1_{\bbeta\neq\bzero} + o(1)1_{\bbeta=\bzero}$, and $m\to\infty$ as $n\to\infty$. Then $m^{1/2}(\hat{\theta}_m^*-\hat{\theta}_n)$ converges to the same limiting distribution as $n^{1/2}(\hat{\theta}_n-\theta_0)$ conditionally (on the data), in probability.
	\label{thm:bootstrap1}
\end{theorem}

A key challenge in $m$-out-of-$n$ bootstrap is how to choose $m$. \citet*{bickel2008} proposed an adaptive choice of $m$ for constructing confidence intervals for extrema, and 
proved that the chosen $m$ satisfies the conditions in Theorem \ref{thm:bootstrap1}.
 In our simulation studies, we have found that their approach, when applied to our setting, did not achieve sufficient power, even though it successfully controlled the type I error rate at the nominal level.
Alternatively, in the context of Q-learning for estimating optimal dynamic treatment regimes, \citet*{chakraborty2013} developed a scheme for selecting $m$ for inference of stage-1 regression parameters that adapts to the degree of non-regularity. The tuning parameter involved in the procedure was chosen by double bootstrap, which is very time consuming in our setting.

We extend Bickel and Sakov's method for choosing $m$ by adding a crude pre-testing step with the goal of improving the power without inflating the type I error rate. 
In particular, we define
\begin{align}
\hat r = 1\{|\sqrt n T_n| < \max (\sqrt{c\log n}, \mbox{ upper } \alpha/(2p) \mbox{-quantile of } N(0,1))\},
\label{eqn:r}
\end{align} where $T_n=\hat{\theta}_n/\hat{\sigma}_n$ is the conventional t-statistic based on the selected covariate $X_{\hat k_n}$, $\alpha$ is the level of significance, and $c>0$ is a tuning parameter. Since $\sqrt n T_n = O_p(1)$ under $H_0$ and $\sqrt n T_n = O_p(\sqrt n)$ under $H_a$, $\sqrt{c\log n}$ on the right hand side of (\ref{eqn:r}) guarantees that  $\hat r \stackrel{P}{\to} 1_{\theta_0=0}$. 
The second component within $\max$ on the right hand side of (\ref{eqn:r}) is used to control the type I error rate in case of small samples (see \citealt{mckeague2015}). Note that $\sqrt{c\log n}$ is an n-term, and the second term in (\ref{eqn:r})  is a p-term. Intuitively,
when p is large, we expect that p plays a more important role than n, and thus p-term should
dominate the n-term, and vice-versa. The tuning parameter $c$ controls the balance of these two terms. We find  $c=2$ works well in various simulation settings, and suggest to use $c=2$. 
If $\hat r=0$, we consider it as a crude rejection of $H_0$, and propose to use the regular $n$-out-of-$n$ bootstrap to conduct a refined test. On the other hand, $\hat r = 1$ indicates that there might be some non-regularity, and we propose to use Bickel and Sakov's method for choosing $m$.  The complete algorithm is given below.
\begin{enumerate}
	\item Calculate $\hat r$ defined in (\ref{eqn:r}). If $\hat r=0$, then choose $\hat m = n$.  Otherwise, continue with steps 2-4 to obtain Bickel and Sakov's estimate $\hat m^{BS}$.

	\item  Consider a sequence of $m$'s of the form: $m_j = \lceil d^j n\rceil$, for $j = 0,1,2,..$ and $0 < d < 1$, where $\lceil x\rceil$ denotes the smallest integer $\ge x$, and $d$ is a  tuning parameter.

	\item For a given data set (with estimate $\hat{\theta}_n$), and for all $j$, define the bootstrap empirical distribution function:
	$R_{m_j}^B(t,\hat{\theta}_n) = \sum_{b=1}^B1_{m_j^{1/2}(\hat{\theta}_{m_j}^{*,b}-\hat{\theta}_n)\leq t}/B, 
$
	where $\hat{\theta}_{m_j}^{*,b}$ is the $m_j$-out-of-$n$ bootstrap version of the estimate $\hat{\theta}_n$ from the $b$-th bootstrap sample, $b=1,\ldots, B$.

	\item Following \citet*{bickel2008},  set $m$ as the minimizer of the sup-norm of the successive differences between the bootstrap empirical distribution functions:
	$\hat m^{BS}=\arg\min_{m_j}\sup_t|R_{m_j}^B(t,\hat{\theta}_n)-R_{m_{j+1}}^B(t,\hat{\theta}_n)|.$

	\item Output $\hat m=(1-\hat r)n + \hat r\hat m^{BS}.$
\end{enumerate}

The tuning parameter $d$ in Step 2 above can be viewed as a step size since $m_{j+1}/m_j\approx d$.  \citet*{bickel2008} used $d = 0.75$ in their simulation study, and reported robustness to other values. In our setting, we find that our method is pretty robust to the choice of $d\in[0.7, 0.9]$ (see Section S5.4 in the Supplementary Materials). In the simulation, we use $d=0.8$ for our analysis.

\setcounter{equation}{0}
\section{Conditional Sequential Test for Randomized Trials} \label{sec:sequential}
In the previous section, we have proposed a marginal screening test to detect whether there is any covariate that interacts with treatment. 
If  the null in (\ref{eqn:h0theta}) is rejected, we will select $\hat k_n$ as the most informative predictor of the causal treatment effect. In this section, we extend our test to  detect additional treatment-by-covariate interactions.

The procedure is carried out in a forward stepwise fashion. At each step, 
let $J\subset\{1,\ldots,p\}$ denote the index set such that $\{X_j: j\in J\}$ have been identified as having significant interaction with treatment in previous steps. We aim to test whether there is any $X_k\in \{X_j : j\in J^C\}$ that interacts with treatment. 
Specifically, we re-write model (\ref{eqn:simplemodel}) as
\begin{align}
Y=h_0(\bX) + (\alpha_0 + \bX^\T_J \bbeta_{0,J}+\bX^\T_{J^C} \bbeta_{0,J^C})A+\epsilon
\label{eqn:model1}
\end{align}
where $\bX_J = \{X_j: j\in J\}$ and $\bX_{J^C} = \{X_j: j\in J^C\}$.
The goal here is to test $\bbeta_{0,J^C}=\bzero$. 

Note that the index set $J$ includes previously selected covariates. Thus the 
the null hypothesis is actually a function of the observed data. This makes any sequential test beyond the initial step a  conditional hypothesis test. Nonetheless, under the alternative hypothesis, the selected covariate is the truly most informative predictor at each step with probability tending to $1$ (see proofs of Theorems \ref{thm:simple} and \ref{thm:sigtest}). So the index set $J$ will converge to a fixed set if the alternative hypotheses in previous steps are true.

For each $k\in J^C$, let $U_k = X_k - \tX_J^\T\bgamma_k$, where $\tX_J = (1, \bX^\T_J)^\T$ and $\bgamma_k = \arg\min_{\bgamma}\left\{E\big[W(X_k - \tX_J^\T\bgamma)\big]^2\right\}$. That is, $\widetilde\bX_J^\T\bgamma_k$ is the weighted projection of $X_k$ on the space spanned by 
$\widetilde\bX_{J}$. 
After algebraic simplification, we can reformulate model (\ref{eqn:model1}) as
\begin{align}
Y = h'_0(\bX) + (\alpha_0' + \bU^\T\bbeta_{0,J^C})A + \epsilon', \label{eqn:model2}
\end{align}
where $\bU=\{U_k: k\in J^C\}$; meanwhile
\begin{align*}
& h_0'(\bX) =  h_0(\bX) + q_0(\bX)\left[\tX_J - \frac{E(W^2\tX_J)}{EW^2}\right]^\T\blambda,\\
\alpha_0' = &\, \frac{E(W^2\tX_J^\T)}{EW^2}\blambda,
\mbox{ and } \epsilon' =  \epsilon + W\left[\tX_J - \frac{E(W^2\tX_J)}{EW^2}\right]^\T\blambda,
\end{align*}
where $\blambda = (\alpha_0,\bbeta^\T_{0,J})^\T+ \bGamma\bbeta_{0,J^C}$ 
with $\bGamma$ being a parameter matrix with columns consisting of $\{\bgamma_k: k\in J^C\}$.

Note that (\ref{eqn:model2}) is of similar form as model (\ref{eqn:simplemodel}), with $\bX$ replaced by $\bU$. 
Thus to test $\bbeta_{0,J^C}=0$ is equivalent to testing
$H_0: \theta_0' = 0 \mbox{ vs. } H_a: \theta_0'\neq 0,$
where 
\begin{align*}
\theta_0' = &\frac{E[W\big(U_{k_0'}-E(W^2U_{k_0'})/EW^2\big)(Y-E(Y|\bX))]}{E\Big\{\Big[W\Big(U_{k_0'} - E(W^2U_{k_0'})/EW^2\Big)\Big]^2\Big\}}
= \frac{\Cov(WU_{k_0'}, W\bU^\T\bbeta_{0, J^C})}{E\Big(W^2U^2_{k_0'}\Big)},\\
\mbox{and } k_0' = & \arg\max_{k: k\in J^C} \big|\Corr(WU_k, W\bU^\T)\bbeta_{0, J^C}\big|.
\end{align*}

Similar to the approach as described in Section \ref{sec:marginal},  we can estimate $\theta_0'$  by $\hat{\theta}_n'$ using empirical quantities. 
The theorem below gives the asymptotic distribution of $\hat{\theta}'_n$ and establishes the bootstrap consistency.
\begin{theorem}
	\label{thm:sigtest} Assume conditions i) - iii) in Theorem  \ref{thm:simple} hold. Suppose $k_0' = \arg\max_{k: k\in J^C} \big|\Corr(WU_k, W\bU^\T\bbeta_{0, J^C})\big|$ is unique when $\bbeta_{0,J^C}\neq \bzero$. Then under model  (\ref{eqn:model1}),
	$$n^{1/2}(\hat\theta'_n-\theta'_0)\stackrel{d}{\to} \frac{Z_{k_0}}{E(WU_{k_0})^2}1_{\bbeta_{0, J^C}\neq \boldsymbol 0} + \frac{Z_{K}}{E(WU_K)^2}1_{\bbeta_{0, J^C} =  \boldsymbol 0},$$
	where $K=\arg\max_{k in J^C}Z_k^2/E(WU_K)^2$,
	and  $\{Z_k: k\in J^C\}$ is a mean-zero normal random vector with covariance matrix
	$\Sigma$ given by that of the random vector with components
	\begin{align*}
	WU_k\Big\{Y &-\tilde{\phi}(\bX)-\frac{E[WU_k(Y-\tilde{\phi}(\bX))]}{E(WU_k)^2}WU_k \\
	& -W\tX_J^\T \left(EW^2\widetilde{\bX}_J\widetilde{\bX}^\T_J\right)^{-1}E\left[W\tX_J\left(Y-\tilde{\phi}(\bX)\right) \right]\Big\}
	\end{align*}
	for $k\in J^C$, and $\Sigma$ is assumed to exist.
	
	Furthermore, Let $\hat{\theta}_m'^{*}$ be the $m$-out-of-$n$ bootstrap analog of $\hat{\theta}_n'$. Assume $m/n=O(1) 1_{\bbeta_{0,J^C}\neq \bzero} + o(1) 1_{\bbeta_{0,J^C}= \bzero}$, and $m\to\infty$ as $n\to\infty$.  Then $m^{1/2}(\hat{\theta}_m'^{*}-\hat{\theta}_n')$ converges to the same limiting distribution as $n^{1/2}(\hat{\theta}_n'-\theta_0')$ conditionally (on the data), in probability.
\end{theorem}
The test of $\theta_0'=0$ can be conducted using either the sampling from null procedure or the $m$-out-of-$n$ bootstrap procedure as described in Section \ref{sec:marginal}.

\setcounter{equation}{0} 
\section{Extension to Allow for Double Robustness}
\label{sec:observe}

Methods presented in previous sections are designed for scenarios where the propensity score $q_0(\bX)$ is known. 
In this section, we further improve the procedure to allow for {\em double robustness} in the sense that as long as $q_0(\bX)$ or $h_0(\bX)$ is consistently estimated, a valid inferential procedure as described previously can be established. This would be particularly useful in observational studies, where the propensity score $q_0(\bX)$ is usually unknown.

We start with model (\ref{eqn:model1}), where the goal is to test $\bbeta_{0,J^C} = \bzero$ after $\{X_j: j\in J\}$ have been detected in previous steps. Note that the initial test of $\bbeta_0=\bzero$ is a special case with $J = \emptyset$. 
Let $\hat q_n(\bX)$ and $\hat h_n(\bX)$ be the estimates of $q_0(\bX)$ and $h_0(\bX)$ based on the data, and $\tilde q(\bX)$ and $\tilde h(\bX)$ be the limit of $\hat q_n(\bX)$ and $\hat h_n(\bX)$, respectively (see Appendix \ref{sec:appendix} for conditions on $\tilde q(\bX)$ and $\tilde h(\bX)$). 
Denote $\tW := A - \tilde q(\bX)$. 

To ensure double robustness, parameter estimates are often obtained through solving estimating equations, known as G-estimation for structural mean models (\citealt{robins1989, robins1994}).
Past research in this area focused on fitting a full causal treatment effect model restricted to a small set of variables, and studying the efficiency of the estimate. In this section, we apply g-estimation marginally on each covariate and conduct test based on the selected most informative covariate.
Specifically, 
for each $k\in J^C$, let $(\bdelta_k,\psi_k)$ be the solution to 
\begin{align}
E\Big\{(\tX_J^\T,X_k)^\T\tW\big[Y-\tilde h(\bX)-(\tX_J^\T\bdelta + X_k\psi)A\big]\Big\}=0.
\label{eqn:OBSequation}
\end{align}
This yields
$\psi_k =   \left[E(A\tW L^2_k)\right]^{-1}E[\tW(Y-\tilde h(\bX))L_k],$
where $L_k = X_k - \tX^\T_J\boldeta_k$ with $\boldeta_k = \arg\min_{\boldeta}E\big[A\widetilde W(X_k-\tX^\T_J\boldeta)^2\big]$. 

To identify the most informative predictor, we need to identify the optimization objective function corresponding to equation (\ref{eqn:OBSequation}).
Note that under model  (\ref{eqn:model1}), when $\tilde q(\bX)=q_0(\bX)$ or $\tilde h(\bX)=h_0(\bX)$ a.s., the left hand side of (\ref{eqn:OBSequation}) is equivalent to $E\big\{(\tX_J^\T,X_k)^\T A\tW\big[\alpha_0 + \bX^\T\bbeta_0 -(\tX_J^\T\bdelta + X_k\psi)\big]\big\}$, where  the quantity inside the expectation can be viewed as a quasi-likelihood score function. Thus the solution to (\ref{eqn:OBSequation})  satisfies
\begin{align*}
(\bdelta_k,\psi_k) = \arg\min_{(\bdelta,\psi)}E\big[A\tW(\alpha_0 + \bX^\T\bbeta_0 - \tX^\T_J\bdelta - X_k\psi)^2\big].
\end{align*}
Intuitively, $\tX^\T_J\bdelta_k + X_k\psi_k$ can be viewed as the best weighted linear approximation of the causal treatment effect $T(\bX)=\alpha_0 + \bX^\T\bbeta_0$ based on $(\tX_J, X_k)$. Thus it is natural to define the most informative predictor in $\{X_j: j\in J^C\}$ by
$$k^o_0 :=  \arg\min_{k: k\in J^C} E\big[A\tW(\alpha_0 + \bX^\T\bbeta_0 - \tX^\T_J\bdelta_k - X_k\psi_k)^2\big] = \arg\max_{k: k\in J^C} \Big[E\big(A\tW L_k^2\big)\psi_k^2\Big].$$ And the hypothesis of interest is
\begin{align}
H_0: \psi_0 =0 \mbox{ vs. } H_a: \psi_0\neq 0,
\label{eqn:psi_0}
\end{align}
where $\psi_0 := \psi_{k^o_0}$.
We can estimate $\psi_0$ by $\hat{\psi}_n := \hat{\psi}_{\hat k^o_n}$, where
$\hat{\psi}_k =  \bbP_n\Big[\hW (Y-\hat h_n(\bX)) \hat L_k\Big]\Big/ \bbP_n\big(A\hW \hat L_k^2\big)$ 
and  $\hat k^o_n = \arg\max_{k \in J^C}  \Big[\hat{\psi}_k^2 \bbP_n\big(A\hW \hat L_k^2\big)\Big],$ with $\hW = A - \hat q_n(\bX)$, $\hat L_k = X_k - \tX^\T_J\hat\boldeta_k$, and $\hat\boldeta_k = \arg\min_{\boldeta}\bbP_n\Big[A\widehat W(X_k-\tX^\T_J\boldeta)^2\Big]$.

The asymptotic distribution of $\hat{\psi}_n$ depends on the limiting behavior of $\hat q_n(\bX)$ and $\hat h_n(\bX)$, and is difficult to estimate. Thus $m$-out-of-$n$ bootstrap will play an important role in obtaining valid inference for $\psi_0$. Let $\hat{\psi}_m^*$ denote the bootstrap analog of $\hat\psi_n$. Below we give the asymptotic distribution of $\hat{\psi}_n$ and prove the bootstrap consistency. A complete list of assumptions is given in Appendix \ref{sec:appendix}.
\begin{theorem}
	\label{thm:observe}
	Suppose Assumptions (A\ref{cond:moment1})-(A\ref{cond:dr}) in Appendix \ref{sec:appendix} hold. Then under model  (\ref{eqn:model1}),
	$$n^{1/2}(\hat\psi_n-\psi_0)\stackrel{d}{\to} \frac{\tilde Z_{k^o_0}}{E(A\tW L_{k^o_0}^2)}1_{\bbeta_{0, J^C}\neq \boldsymbol 0} + \frac{\tilde Z_{K}}{E(A\tW L_K^2)}1_{\bbeta_{0, J^C} =  \boldsymbol 0},$$
	where $K=\arg\max_{k \in J^C} \big[\tilde Z_k^2/E(A\tW L_k^2)\big]$, and 
	$\{\tilde Z_k: k\in J^C\}$ is a normal random vector defined in 	(\ref{eqn:tZk}).

	Furthermore, suppose Assumption (A\ref{cond:bs}) in Appendix \ref{sec:appendix} holds, $m/n = O(1) 1_{\bbeta_{0, J^C} \neq  \boldsymbol 0} + o(1) 1_{\bbeta_{0, J^C} =  \boldsymbol 0}$ and $m\to\infty$ as $n\to\infty$.  Then under model (\ref{eqn:model1}), $m^{1/2}(\hat{\psi}_m^*-\hat{\psi}_n)$ converges to the same limiting distribution as $n^{1/2}(\hat{\psi}_n-\psi_0)$ conditionally (on the data), in probability.
\end{theorem}

Note that, although the doubly robust method can be used for randomized trials, it may cause extra dispersion in variance in our setting. This issue is discussed in Section S4 of the Supplementary Materials and via simulations. 

\setcounter{equation}{0} 
\section{Numerical studies}
\label{sec:numerical}

In this section, we study the performance of the proposed sequential testing procedure using simulated data, and give illustrations of the approach in two real data examples. 

\subsection{Simulations} 

Below we briefly summarize simulations studies. See Section S5 in the Supplementary Materials for the details of simulations.

In the randomized trial setting, we compare the proposed sampling from null (NULL), $m$-out-of-$n$ bootstrap ($\hat m$-boot), and the doubly robust $m$-out-of-$n$ bootstrap ($\hat m$-boot-DR) procedures with four competing methods: Likelihood ratio test (LRT), Multiple testing with Bonferroni correction (BONF), $n$-out-of-$n$ bootstrap ($n$-boot), and $m$-out-of-$n$ bootstrap with $m$ chosen by Bickel and Sakov's method ( $\hat m^{BS}$-boot). 
Three data generating models are considered: i) null model; ii) model with one active interaction term; and iii) model with two equally active interaction terms. The sequential testing procedure is carried out to evaluate the power (when there is at least one active predictor in the candidate set) or type I error rate (when there is no active predictor remaining in the candidate set)  at each step.
The two proposed methods for randomized trials (NULL and $\hat m$-boot) provide good control of type I error rate and good power in all cases. $\hat m$-boot-DR, $\hat m^{BS}$-boot and LRT are less powerful as compared to NULL and $\hat m$-boot.  $n$-boot fails to control the type I error rate. When the components of $\bX$ are uncorrelated, BONF is as good as our proposed methods in terms of type I error rate control and power. However, when the components of $\bX$ are highly correlated, BONF is less powerful for large $p$ (see Tables S1 and S2 in the Supplementary Materials). 
  
In the observational study setting, we compare the $\hat m$-boot-DR method with  $\hat m^{BS}$-boot and $n$-boot methods.
We consider four data generating models: two with correctly specified $q_0(\bX)$ and mis-specified $h_0(\bX)$, two with mis-specified  $q_0(\bX)$ and correctly specified $h_0(\bX)$. Linear logistic regression model with adaptive lasso is used to estimate the propensity score model $q_0(\bX)$, and linear regression with adaptive lasso is used to estimate the main effect $h_0(\bX)$. The proposed $\hat m$-boot method provides good control of type I error rate and good power in all cases. $\hat m^{BS}$-boot lacks  power as compared to $\hat m$-boot, and $n$-boot fails to control the type I error rate (see Table S3 in the Supplementary Materials). 

We also compare our methods with the kernel based method proposed by \cite{shen2016} (KM$_l$) and the GESAT method proposed by \cite{lin2013test} for the global test of no treatment-by-covariate interactions. Both KM$_l$ and GESAT are designed to test for the integrated  effect of all covariates, with KM$_l$ based on randomized trials and GESAT based on correct specification of both main effect of covariates and interaction effects. As a result, we expect that KM$_l$ and GESAT are more powerful in the case of weak dense signals while our method works better in the case of strong sparse signals. This is indeed the case (see Table S4 in the Supplementary Materials). All methods perform better when covariates are correlated. In addition, KM$_l$ fails to control type I error rate in observational studies and GESAT fails when the main effect is misspecified (see Table S5 in the Supplementary Materials).

\subsection{Nefazodone-CBASP trial example}    	

The Nefazodone-CBASP trial was conducted to compare the efficacy of  three treatments for chronic depression (\citealt{keller2000}). In this trial, $681$ patients were randomly assigned to $12$ weeks of outpatient treatment with nefazodone, the cognitive behavioral-analysis system of psychotherapy (CBASP), or the combination of the two treatments. Various assessments were taken throughout the study, among which the score on the 24-item Hamilton Rating Scale for Depression (HRSD) was the primary outcome. Low HRSD scores are desirable. Primary analysis showed that, on average, the combination treatment was significantly better than any single treatment (p-value $< 0.001$ for both comparisons), and there was no overall difference between two single treatments.

In the current analysis, we conduct two comparisons: combination vs nefazodone alone and combination vs. CBASP alone, to see if there is any covariate-by-treatment interactions. The outcome $Y$ is reduction in 24-item HRSD score from baseline, and there are $50$ baseline covariates. We consider $n=656$ patients for which the final 24-item HRSD were observed. In both comparisons, we estimate $\tilde \phi(\bX)=E(Y|\bX)$ by ridge regressions, and carry out the sequential tests in 5-steps. The selected covariates and p-values from different methods are presented in Table \ref{table:CBASP}.

For comparison of combination treatment vs. CBASP alone (n=438), the regular $n$-boot suggests three important covariates: Subthreshold panic disorder ($p=0.016$), Psychotherapy for past depression ($p=0.034$), and Alcohol abuse ($p=0.050$). However, this is not supported by any other methods. 
For comparison of combination treatment vs Nefazodone alone ($n=440$), again regular $n$-boot identifies three covariates:  Psychotherapy for past depression ($p<0.001$), Alcohol dependence ($p=0.020$), and Obsessive compulsive disorder ($p=0.030$). With LRT, all p-values are greater than $0.1$. In between, the NULL, $\hat m$-boot and BONF methods show that only ``Psychotherapy for past depression" has a p-value less than $0.1$, which indicates that this may be worth further investigation. 

\begin{table}[h!]
	\caption{p-values of sequentially selected covariates in the Nefazodone-CBASP trial example. (PD: Panic Disorder; PsyPD: Psychotherapy for Past Depression; GAD: Generalized Anxiety Disorder; MDD: Major Depression Disorder; OCD: Obsessive Compulsive Disorder; AD-NOS: Anxiety Disorder not otherwise specified.) }
\label{table:CBASP}\par
\vskip .2cm
\centerline{\tabcolsep=3truept
	\begin{tabular}{llccccc}
			\hline
		Steps &	Covariates selected & NULL & $\hat m$-boot & n-boot & LRT & BONF \\
		\hline
		\multicolumn{2}{l}{Combination vs CBASP} & & & &\\
		1& Subthreshold PD& 0.628 & 0.578 & 0.016 & 0.792   &  1.000\\
		2&PsyPD& 0.622 & 0.754 & 0.034 & 0.903  &  1.000\\
		3&Alcohol abuse & 0.764 & 0.302 & 0.050 & 0.966   &  1.000\\
		4&Threshold GAD & 0.983 & 0.946 & 0.832 & 0.990   & 1.000\\
		5&moderate MDD & 0.993 & 0.564 & 0.622 & 0.994   & 1.000\\ 
		\cline{2-7}
		\multicolumn{2}{l}{Combination vs Nefazodone} & & & &\\
		1&Past Psychotherapy & 0.096 & 0.088 & 0.000 & 0.147 & 0.086\\
		2&Alcohol dependence & 0.369 & 0.122 & 0.020 & 0.386 & 0.458\\
		3&OCD & 0.562 & 0.758 & 0.030 & 0.604 & 0.862\\
		4&AD-NOS & 0.907 & 0.296 & 0.122 & 0.774 & 1.000\\
		5&Atypical MDD & 0.925 & 0.564 & 0.140 & 0.852 & 1.000\\ 
			\hline
	\end{tabular}}

\end{table}

\subsection{COPES and CODIACS example}
The COPES and CODIACS studies were conducted to compare the stepped care approach to standard care for patients with post-ACS (acute coronary syndrome) depression  (\citealt{COPES, CODIACS}). Each trial enrolled about 150 patients. In both trials, patients were randomly assigned to 6 months of stepped care or usual care.	
Stepped Care participants were assigned to psychotherapy and/or antidepressant based on their preferences and a team of clinicians' recommendations.  Usual care participants received psychotherapy and/or antidepressant from their current physicians. Depressive symptom changes, assessed with the Beck Depression Inventory (BDI) score, was the primary outcome for CODIACS and a secondary outcome for COPES.

As an illustrative example, for the combined data from the above two studies, we consider treatment $A$ as whether a patient had received psychotherapy more than half of the time during the 6-month study period. Note that while the original two studies were randomized trials, the ``treatment received'' variable $A$ as defined above is observational in nature. There are $26$ baseline covariates, including patient demographics, symptoms, and severity of symptoms in different domains (cardiac, depression, etc). The outcome is reduction in BDI at 6 months from baseline. The sample size $n=265$ for which the final BDI and treatment information are available.

For the propensity score model, after initial analysis we find that the treatment $A$ as defined above strongly depends on the treatment arm and patient preference. So we estimate the propensity score by logistic regression with treatment preference and whether the patient was assigned to stepped care or usual care. The main effect $h_0(\bX)=E(Y|\bX, A=0)$ is estimated using ridge regression with 26 baseline covariates using patients in the $A=0$ group. We conduct 5 steps sequential tests, and the results from $\hat m$-boot and $n$-boot are presented in Table \ref{table:CODIACS}. The $n$-boot identifies two important treatment-by-covariate interactions: NEMC Role limitation emotional T-score ($p=0.004$) and GRACE score ($p=0.024$), while our proposed $\hat m^{ad}$-boot method suggests that only NEMC Role limitation emotional T-score may be worth further investigation.

\begin{table}[h!]
		\caption{p-values of sequentially selected covariates in the COPES and CODIACS  example.}
	\label{table:CODIACS}\par
	\vskip .2cm
\centerline{\tabcolsep=3truept	\begin{tabular}{llcc}
	\hline
		Steps &	Covariates selected & $\hat m$-boot-DR & n-boot  \\
		\hline
		1& NEMC Role limitation emotional T-score  & 0.028 & 0.004 \\
		2&GRACE score  & 0.330 & 0.024 \\
		3&NEMC General health T-score & 0.504 & 0.112 \\
		4&BDI$\ge 29$ & 0.672 & 0.186 \\
		5&Prefer psychotherapy only & 0.690 & 0.232 \\ 
		\hline
	\end{tabular}}

\end{table}

\setcounter{equation}{0} 

\section{Discussion}  \label{sec:discussion}         			

This paper develops a novel inference procedure to sequentially identify treatment-by-covariate interactions based on data from randomized trials as well as observational studies. The proposed method guarantees rigorous control of type I error rate at each step, and has greater power than competing testing procedures. 
Although derivation of the asymptotic results assumes fixed dimension $p$,  numerical studies evidenced that the proposed test continues to work when $p$ is large. Theoretical investigation of the diverging $p$ case is a challenging and interesting topic for future research.

Another challenge is to provide theoretical support to effectively control the false positive rate over the whole sequence
of forward stepwise path. As discussed in  \citet{tibshirani2016}, sequential testing is typically only validated at each step. \citet{gsell2016} proposed stopping rules for exact control of ordered false discovery rate control under the assumption that the null p-values are independent. It would be worthwhile to study how to extend their methods to our setting so that 
the stepwise guarantees can be converted into stopping rules with desired inferential properties.


In this paper we consider the case of binary treatments. When there are more than two treatments (e.g. $L$ arms), the interaction effect between treatment and a covariate includes $L-1$ terms, and a test statistic can be built based on the sum of square of the $L-1$ standardized parameter estimates. This would be an interesting extension for future research.

We have restricted attention to a single-stage treatment-by-covariate interaction problem. However,
time-varying treatments are common, and are needed, e.g., for individuals with
a chronic disease who experience a waxing and waning course of illness. The
goal then is to identify informative covariates of treatment effect at each stage. Q-learning and A-learning are extensions of regression to multi-stage setting (\citealt*{murphy2003, murphy2005b, moodie2007}; \citealt{schulte}). It is well known that the regression coefficients for variables at stages prior to the last are non-regular (\citealt*{robins2004, moodie2010, chakraborty2010,  chakraborty2013}; \citealt{laber2014a, song2015}). In that case, the selection procedure adds another layer of non-regularity to this already non-regular problem.
It would be interesting albeit challenging to
extend our approach to the multi-stage setting. We view this as an important future work.

\vskip 14pt
\noindent {\large\bf Supplementary Materials}

The online supplementary materials contain proofs of the theorems, extension of Theorem \ref{thm:simple} to non-unique $k_0$, discussion of the doubly robust method  in randomized trials, and details of simulation studies.
\par
\vskip 14pt
\noindent {\large\bf Acknowledgements}

Min Qian's research is supported in part by  NIH Grant R21MH108999, 2R01GM095722, and R01MH109496.
Bibhas Chakraborty and Raju Maiti's research are supported in part by MOE2015-T2-2-056 grant from Singapore's Ministry of Education and the start-up grant from Duke-NUS Medical School. Ying-Kuen Cheung's research is supported in part by  NIH Grant R21MH108999 and R01MH109496.
\par
\setcounter{equation}{0} 
\appendix
\section{Assumptions of Theorems \ref{thm:observe}}
\label{sec:appendix}
Theorem \ref{thm:observe} requires the following assumptions.
\renewcommand{\labelenumi}{(A\arabic{enumi})}
\begin{enumerate}
	\item $EX_k^4 <\infty$ for $k=1,\ldots, p$. \label{cond:moment1}
	\vspace{-0.1in}
	\item  \label{cond:normal} There exist functions $\tilde h(\bX)$ and $\tilde q(\bX)$ such that  $n^{1/2}[\hat h_n(\bx) - \tilde h(\bx)] = \Delta_h(\bx)\hat S_h + o_P(1)$ and $n^{1/2}[\hat q_n(\bx) - \tilde q(\bx)] = \Delta_q(\bx)\hat S_q + o_P(1)$, where $\Delta_h(\bx)$ and $\Delta_q(\bx)$ are vector-valued deterministic functions of $\bx$, and
	$\hat S_h$ and $\hat S_q$ are data dependent random vectors satisfying
	
	i). $\Delta_h(\bX)$ and $\Delta_q(\bX)$ are square integrable random vectors; and
	\begin{align*}
	\mbox{ii) } & \Bigg(\Big\{\bbG_n\Big[\tW L_k\Big(Y-\tilde h(\bX)- \psi_k A L_k-E\Big[\tW \Big(Y-\tilde h(\bX)\Big)\tX_J^\T\Big]
	\\
	& \quad\quad\quad\quad\quad\quad\quad\quad \times \Big[E \Big(A\tW \tX_J\tX_J^\T\Big)\Big]^{-1} A\tX_J\Big)\Big]\Big\}_{k\in J^C}, 
	\hat S_h, \, \hat S_q \Bigg)^\T \\
	& \quad\quad\quad\quad \stackrel{d}{\to} (\{Z^o_k: k\in J^C\}, S_h, S_q)^\T \sim N(0, \Sigma^o)
	\end{align*}
	for some variance-covariance matrix $\Sigma^o$ assumed to exist.
		\vspace{-0.1in}
	\item  The error term $\epsilon$ in model (\ref{eqn:simplemodel}) has mean zero, finite variance, and is uncorrelated with $(\tW, \tW\bX)$, where $\tW = A - \tilde q(\bX)$. \label{cond:err1}
		\vspace{-0.1in}
	\item $k_0^o$ is unique when $\bbeta_{0, J^C}\neq 0$. \label{cond:unique1}
		\vspace{-0.1in}
	\item \label{cond:dr} $\tilde q(\bX)=q_0(\bX)$ or $\tilde h(\bX)=h_0(\bX)$ a.s.
		\vspace{-0.1in}
	\item \label{cond:bs} Let $\hat q_m^*(\bX)$ and $\hat h_m^*(\bX)$ be estimates of $q_0(\bX)$ and $h_0(\bX)$ based on the bootstrap sample of size $m$. Assume $m^{1/2}[\hat h_m^*(\bx) - \hat h_n(\bx)] = \Delta_h(\bx)\hat S^*_h + o_{P_M}(1)$ and $m^{1/2}[\hat q_m^*(\bx) - \hat q_n(\bx)] = \Delta_q(\bx)\hat S^*_q + o_{P_M}(1)$ conditionally on the data (in probability), where $\Delta_h(\bx)$ and $\Delta_q(\bx)$ are defined in Assumption (A\ref{cond:normal}), and
	$\hat S^*_h$ and $\hat S^*_q$ are bootstrap sample dependent random vectors satisfying
	\begin{align*}
	& \Bigg(\Big\{\bbG_m^*\Big[\tW L_k\Big(Y-\tilde h(\bX)- \psi_k A L_k-E\Big[\tW \Big(Y-\tilde h(\bX)\Big)\tX_J^\T\Big]\\
	&\quad\quad\quad\quad\quad\quad\quad\quad \times\Big[E \Big(A\tW \tX_J\tX_J^\T\Big)\Big]^{-1}A\tX_J\Big)\Big]\Big\}_{k\in J^C},
	\hat S^*_h, \, \hat S^*_q \Bigg)^\T \\
	&\quad\quad \stackrel{d}{\to} (\{Z^o_k: k\in J^C\}, S_h, S_q)^\T \sim N(0, \Sigma^o)
	\mbox{  conditionally, in probability.}
	\end{align*}
\end{enumerate}
\noindent\textbf{Remark 1.} $\{\tilde Z_k: k\in J^C\}$ in Theorem \ref{thm:observe} is  defined as, for $k\in J^C$,
\begin{align}
\tilde Z_k & = Z_k^o -E\Big[\tW L_k\Delta_h(\bX)\Big]S_h 
- E\Bigg\{L_k \Big(Y-\tilde h(\bX) - \psi_k A L_k \nonumber\\
& -E\Big[\tW \Big(Y-\tilde h(\bX)\Big)\tX_J^\T\Big]
\Big[E \Big(A\tW \tX_J\tX_J^\T\Big)\Big]^{-1}A\tX_J\Big) \Delta_q(\bX)\Bigg\}S_q  
\label{eqn:tZk}
\end{align}

\noindent\textbf{Remark 2.} Assumptions (A\ref{cond:normal}) and (A\ref{cond:bs}) require that the original sample and bootstrap sample estimates of $h_0(\bx)$ and $q_0(\bx)$ are well behaved. One can verify that, under appropriate conditions, the assumptions hold when $h_0(\bx)$ and $q_0(\bx)$ are estimated using linear/logistic regression, ridge regression, or variable selection methods with oracle properties (e.g. SCAD (\citealt{fan2001}), adaptive Lasso (\citealt{zou2005})).

\par

\markboth{\hfill{\footnotesize\rm MIN QIAN, BIBHAS CHAKRABORTY， RAJU MAITI， YING KUEN CHEUNG} \hfill}
{\hfill {\footnotesize\rm SEQUENTIAL TEST OF INTERACTION EFFECTS} \hfill}

\bibhang=1.7pc
\bibsep=2pt
\fontsize{9}{14pt plus.8pt minus .6pt}\selectfont
\renewcommand\bibname{\large \bf References}

\expandafter\ifx\csname
natexlab\endcsname\relax\def\natexlab#1{#1}\fi
\expandafter\ifx\csname url\endcsname\relax
  \def\url#1{\texttt{#1}}\fi
\expandafter\ifx\csname urlprefix\endcsname\relax\def\urlprefix{URL}\fi

\bibliographystyle{dcu}
\bibliography{reference}

\lhead[\footnotesize\thepage\fancyplain{}\leftmark]{}\rhead[]{\fancyplain{}
	\rightmark\footnotesize{} }
%
%
%
%
%

\vskip .65cm
\noindent
Department of Biostatistics, Columbia University, 722 West 168th St., NYC, NY 10032, USA.
\vskip 2pt
\noindent
E-mail: mq2158@cumc.columbia.edu; \quad
Phone: (212) 305-6448;\quad
 Fax: (212) 305-9408
\vskip 2pt

\noindent
Centre for Quantitative Medicine, Duke-NUS Medical School,
Singapore 169857.
\vskip 2pt
\noindent
E-mail: bibhas.chakraborty@duke-nus.edu.sg
\vskip 2pt

\noindent
Centre for Quantitative Medicine, Duke-NUS Medical School, Singapore 169857.
\vskip 2pt
\noindent
E-mail: raju.maiti@duke-nus.edu.sg

\vskip 2pt

\noindent
Department of Biostatistics, Columbia University, 722 West 168th St., NYC, NY 10032, USA.
\vskip 2pt
\noindent
E-mail: yc632@cumc.columbia.edu
\end{document}